\documentclass[11pt]{article}

\usepackage[margin=1in]{geometry}
\usepackage[T1]{fontenc}
\usepackage[protrusion=false]{microtype}
\usepackage{amsmath}
\usepackage{amssymb}
\usepackage{mathtools}
\allowdisplaybreaks

\usepackage{graphicx}
\usepackage{caption}
\usepackage{float}
\usepackage{dcolumn}
\usepackage{tabulary}
\usepackage{multirow}
\usepackage{makecell}
\usepackage{array}
\usepackage{hhline}

\usepackage[authoryear,round]{natbib}
\usepackage{etoolbox}

\AtBeginEnvironment{thebibliography}{%
  \small
  \setlength{\emergencystretch}{4em}%
  \sloppy
}

\usepackage{xcolor}
\usepackage[normalem]{ulem}

\newif\ifrevision
\revisionfalse

\newcommand{\rev}[1]{%
  \ifrevision{\color{blue}#1}\else#1\fi
}

\usepackage[
  colorlinks=true,
  linkcolor=blue,
  citecolor=blue,
  urlcolor=blue
]{hyperref}

\title{Propagating fronts of convection rolls in
Rayleigh--B\'enard convection}

\author{%
Saikat Mukherjee$^{1,*}$ and Mark R. Paul$^{2}$\\[0.6em]
\small $^{1}$Department of Mechanical Engineering,
Iowa State University, Ames, Iowa 50011, USA\\
\small $^{2}$Department of Mechanical Engineering,
Virginia Tech, Blacksburg, Virginia 24061, USA\\[0.4em]
\small $^{*}$Corresponding author:
\href{mailto:saikatm@iastate.edu}{saikatm@iastate.edu}
}

\date{}

\begin{document}

\maketitle

\begin{center}
\small
Published in Mukherjee, S., \& Paul, M. R. (2026), \textit{Journal of Fluid Mechanics}, 1043, A44. \href{https://doi.org/10.1017/jfm.2026.12113}{doi:10.1017/jfm.2026.12113}
\end{center}

\begin{abstract}
 We investigate the propagation of counter-rotating convection rolls in Rayleigh--B\'enard convection initiated locally in a quiescent fluid layer under supercritical conditions. The velocity of the front separating quiescent fluid from the forming convection rolls, and the wavenumber of the convection rolls remaining behind the front, are explored. We numerically investigate fronts of forming convection rolls over five orders of magnitude of the reduced Rayleigh number, $\epsilon$, in 2D and 3D domains, for a broad range of boundary conditions, and for different front initiation approaches. In all cases, the front velocity increases as $\epsilon^{1/2}$ with increasing $\epsilon$ for $\epsilon \lesssim 1$ in agreement with predictions using the amplitude equation. The amplitude equation description of the front velocity remains accurate for $\epsilon \lesssim 10$ except when the Prandtl number is large which yields a velocity that is faster than predicted for a fluid layer far from threshold. The wavenumber of the convection rolls increases linearly with $\epsilon$ in agreement with the wavenumber that maximizes the growth rate of perturbations in the linear regime. Farther from onset, the wavenumber growth transitions to a reduced scaling of $\epsilon^{1/4}$ in agreement with predictions using the Swift--Hohenberg equation in the large $\epsilon$ limit. The scalings describing the wavenumber variation with $\epsilon$ are independent of the domain geometry, boundary conditions, and front initiation method. {However, the front-selected wavenumber at \rev{criticality} does not equal the critical wavenumber of the bulk instability, in general, and depends significantly upon these details.} We compare our results with experimental measurements where possible.
\end{abstract}

\maketitle

\section{Introduction}
Pattern formation in large systems that are driven away from equilibrium is ubiquitous and is  observed in diverse contexts. Important fluid examples include  turbulence~\citep{pope:2000,xiao:2019}, fluid convection~\citep{cross1993pattern,bodenschatz2000recent}, cellular flame fronts~\citep{gi1983instabilities}, chemical reaction fronts~\citep{lee1993pattern,rongy2008dynamics,mukherjee2019velocity,mukherjee2020propagating,mukherjee:2022,gao2023buoyancy}, and droplet freezing~\citep{kant2020pattern}. Pattern formation is also important in many biological systems such as microbial colony organization~\citep{levine1991streaming,tai2022social,chuang2025bacterial}, leaf and flower arrangement in plants~\citep{pennybacker2013phyllotaxis}, wound healing~\citep{sherratt1990models,maini2004travelling}, and neuronal depolarization waves in the brain~\citep{somjen2004ions,dahlem2010two,mukherjee2023quantifying}.

In many of these systems, the patterns emerge behind the propagation of a front separating the patterned and unpatterned regions in space~\citep{ben1985pattern,cross1993pattern,van2003front}. For instance, proliferating band formation of differing cell densities in bacterial colonies subjected to oxygen and nutrient gradients~\citep{chuang2025bacterial}, radially propagating chemical fronts coupled to advection~\citep{maharana2026radially}, cellular clump formation in the wake of ``active'' cell swarms~\citep{ford2025pattern}, sporulation behind expanding biofilm matrix fronts~\citep{srinivasan2018matrix}, and oscillating structures in the wake of bimolecular chemical reactions in fluids influenced by buoyancy and Marangoni instabilities~\citep{budroni2019making,bigaj2023marangoni}. 

The pattern formation that occurs behind a front has been carefully studied in the laboratory using well controlled fluid experiments.  This includes propagating vortex fronts in Taylor--Couette flow~\citep{ahlers1983vortex,ben1985pattern}, vortex fronts in the wake of a cylinder~\citep{yang1989absolute}, patterns behind a front in a layer unstable to the Rayleigh--Taylor instability~\citep{fermigier1992two}, and propagating fronts of vortices~\citep{ben1985pattern,fineberg1987vortex}. The velocity of the front,  and the spatial scales of the fluid structures left in its wake, can depend upon how far the system is driven from equilibrium, the details of the underlying dynamics, the boundary conditions, the mechanism of front initiation, and the size of the domain for finite systems.

Considerable effort has been devoted to building a physical understanding of pattern forming fronts~\citep{dominguez1984marginal,ben1985pattern,dee1983propagating,fineberg1987vortex}. Near critical, an analysis using the amplitude equation yields a prediction for the velocity of the front~\citep{dee1983propagating}. A useful prediction of the selected pattern is often the one which maximizes the growth rate of perturbations in the linear regime~\citep{dominguez1984marginal}. These theoretical predictions have been tested numerically~\citep{lucke1987propagating,kockelkoren2003evidence,ben1985pattern} and experimentally~\citep{fineberg1987vortex}. While the predicted front velocity is generally in good agreement with experimental and numerical results, discrepancies remain in the wavenumber selected by the convection rolls behind the propagating front~\citep{kockelkoren2003evidence,cross1993pattern}.

Rayleigh--B\'enard convection (RBC) is a canonical system for studying pattern formation in a controlled setting that is accessible to experiments~\citep{bodenschatz2000recent} and numerical simulations (cf. \cite{paul2003pattern}). RBC is typically studied as the bulk instability that  results in the buoyancy driven fluid motion of a shallow fluid layer of depth $d$ when heated uniformly from below in a gravitational field. When the temperature difference, $\Delta T\! = T_h \!-\! T_c$, between the hot bottom surface $T_h$ and the cold top surface $T_c$ equals a critical value $\Delta T_c$, yielding the critical Rayleigh number $\mathrm{Ra}_c$, convective motion arises. The fluid motion occurs over the entire fluid layer simultaneously \rev{in the form convection rolls. The convection rolls form} with a critical wavenumber $q_c$ (and critical wavelength $\lambda_c \!=\! 2 \pi/q_c)$.  For an infinite layer of fluid, with no-slip top and bottom surfaces, $\mathrm{Ra}_c\!=\!1707.6$ and $q_c\!=\!3.117$ ($\lambda_c \!=\! 2.016$) where $q_c$ and $\lambda_c$ have been nondimensionalized using $d$~\citep{chandrasekhar:1961}.

In this study, we do not explore the bulk instability where convective fluid motion occurs everywhere in the domain simultaneously. Instead, we consider an initially quiescent layer of fluid which has been locally perturbed to initiate the formation of a convection roll which yields a front propagating into still fluid leaving convection rolls in its wake. The details of our process for accomplishing this are the following.

For time $t\!<\!0$, we set $\Delta T \!=\! 0$ and the entire fluid layer is motionless. At $t\!=\!0$ we prescribe that the temperature of the fluid layer varies linearly from the hot bottom surface to the cold top surface as required by heat conduction in the absence of fluid motion. For $t \! \ge \!0$ we set $\Delta T \!\ge\! \Delta T_c$, the temperature difference then remains at this constant value for all time.  A local perturbation is introduced at $t\!=\!0$ to initiate the formation of a convection roll which results in a propagating front of emerging  convection rolls. \rev{It is useful to note that the timescale for the spontaneous nucleation of convection rolls from the bulk instability scales inversely with the reduced Rayleigh number, near onset~\citep{fineberg1987vortex,ahlers1981amplitude}. However, the front dynamics we study here occur in the window after front initiation and before this bulk instability develops.} Once the bulk instability occurs, the entire fluid layer undergoes convective motion which annihilates the propagating front. Our analysis is of the formation and propagation of the pattern forming front prior to its destruction by the bulk instability.

The front velocity $v_f$, and the wavenumber $q$ of the convection rolls that form behind the front, depend upon the reduced Rayleigh number, $\epsilon$, which is a measure of how far the fluid layer is away from threshold where $\epsilon \!=\! (\mathrm{Ra} \!-\! \mathrm{Ra}_c) / \mathrm{Ra}_c$. As the fluid layer approaches critical from above, $\epsilon \rightarrow0$, the front velocity vanishes and the wavenumber of the forming convection rolls approaches the finite value $q_0$.

Linear stability theory predicts that the front velocity will grow as $v_f \!\sim\! \epsilon^{1/2}$ near threshold~\citep{dee1983propagating}.  \rev{The front-selected wavenumber at $\epsilon\!=\!0$, $q_0$, must be contrasted with the critical wavenumber, $q_c$, which describes the wavenumber of convection rolls that form in an infinite fluid layer due to the bulk fluid instability at $\epsilon=0$.}  Linear stability theory predicts $q_0\!=\!q_c$ for a front of convection rolls in an infinite layer of fluid and for $\epsilon \!\ll\! 1$ the wavenumber increases as $q \!\sim\! \epsilon$~\citep{dominguez1984marginal}.  As we will discuss in detail, $q_0 \!\ne\! q_c$ for all of the conditions we explore.

In a series of pioneering experiments by \cite{fineberg1987vortex} propagating fronts of convection rolls were investigated over a wide range of conditions. These experiments yielded $v_f \!\sim\! \epsilon^{1/2}$ in agreement with predictions using the amplitude equation. However, the measured values of wavenumber were significantly different than the theoretical predictions. The wavelength of the front-selected convection rolls in experiment is described by $\lambda \!=\! \lambda_0 (1-b\sqrt{\epsilon})$ where $\lambda_0 \!=\! 2.29$ ($q_0 \!=\! 2.744$) and $b \!=\! 0.18$. It is important to emphasize that $\lambda_0$ is the wavelength of the front selected convection rolls at $\mathrm{Ra}_c$.

The experimental measurements of the convection rolls disagree with theory in several important respects. The experiments yield $\lambda_0 \!>\! \lambda_c$ where the wavelength of the front-selected convection rolls measured in experiment for $\epsilon \! \rightarrow \! 0$ is 13.6\% larger than $\lambda_c$. In addition, the variation of the wavelength with $\epsilon$ in experiment is different than the theoretical prediction. Expressing the experimental result in terms of wavenumber, expanding for small $\epsilon$, and keeping only the leading term yields $q/q_0 \!=\! 1 \!+\! 0.18 \epsilon^{1/2}$. In summary, theory predicts $q_0\!=\!q_c$ and $q/q_0 \sim \epsilon$ while experimental measurement yields $q_0 \!<\! q_c$ and $q/q_0 \sim \epsilon^{1/2}$.

Previous numerical simulations by \cite{lucke1987propagating} reproduced the theoretically predicted linear scaling $q/q_0 \!\sim \!\epsilon$ with $q_0\!=\!q_c$. The difference between $q_0$ and $q_c$ in experiment was attributed to the details of the apparatus used~\citep{lucke1987propagating,cross1993pattern} and to the slow relaxation of the fronts to their asymptotic values which makes measurements very difficult~\citep{kockelkoren2003evidence}.

In this work, we numerically investigate the propagation of convection rolls over a range of experimentally relevant conditions, including domain size, domain geometry, method of front initiation, and boundary conditions. We explore how these factors influence the front velocity and the wavenumber of the convection rolls. We use a highly flexible spectral element approach~\citep{nek5000} for a range of conditions, including the addition of specific geometrical features of the  experiments of~\cite{fineberg1987vortex}, to quantitatively explore fundamental features of pattern forming fronts.

The remainder of the paper is organized as follows. In Sec.~II, we discuss the general approach, including the governing equations and the details of the computational domains, boundary conditions, and methods of front initiation that are used. In Sec.~III, the numerical results are discussed. We first discuss the propagation of fronts which leave behind straight parallel convection rolls. We begin with a 2D domain and then compare with 3D box domains which include a range of features and boundary conditions. We then discuss fronts of concentric rolls that occur in a large cylindrical domain. Lastly, our conclusions are presented in Sec.~IV.

\section{Approach}

\subsection{Convection Domains}
We use several computational domains to explore the propagation of pattern forming fronts. Our intention with this section is to clearly present the details of the different simulations we conduct prior to discussing the numerical results. We begin our study with a 2D domain which we then extend to become a 3D box domain. With the box domain, we explore a range of experimentally motivated features including different front initiation approaches and the influence of thin fin structures attached to the sidewalls to inhibit fluid motion near the walls. Finally, we explore fronts of concentric rolls formed in a large cylindrical domain. These different cases are all discussed in turn below.

\begin{figure}[h!]
    \begin{center}
        \includegraphics[width=0.8\linewidth]{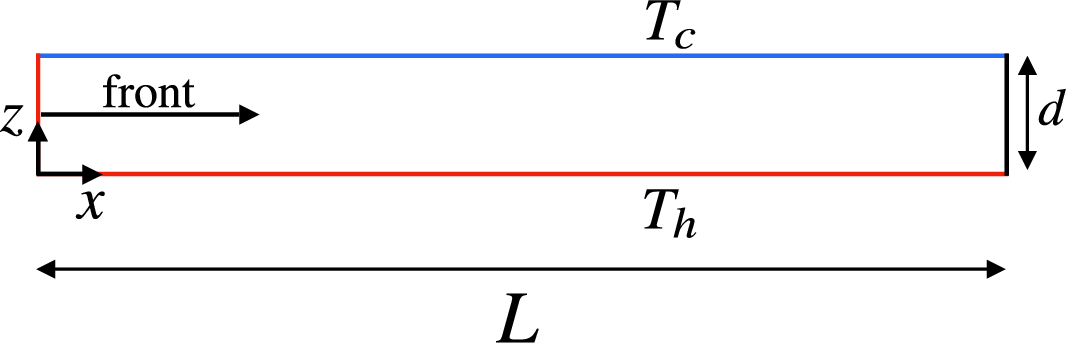}
    \caption{2D domain used to study the propagation of fronts of straight parallel rolls. A fluid layer of length $L$ and depth $d$ with a hot bottom wall at $T_h$ (red) and cold top wall at $T_c$ (blue) where $T_h \!>\! T_c$. The front is initiated using a hot wall located as $x\!=\!0$ (red) and propagates from left to right. The aspect ratios $\Gamma \!=\! L/d$ used in the simulations are  $\Gamma \!=\! 30$ and 60. Gravity opposes the $z$-direction.}
    \label{fig:fig1}
    \end{center}
\end{figure}

The 2D domain, shown in Fig.~\ref{fig:fig1}, has a length $L$ in the $x$-direction where $d$ is the depth of the fluid layer in the $z$-direction. The bottom surface is hot (red) with temperature $T_h$, the top surface is cold (blue) with temperature $T_c$, and gravity opposes the $z$-direction. All material boundaries are no-slip surfaces.  For most of our results, we have used a domain with aspect ratio $\Gamma \!=\! L/d \!=\! 30$. However, we found it necessary to extend the domain to $\Gamma \!=\!60$ to study fronts near onset, $\epsilon \!\ll\! 1$, which required a longer time and larger domain for the fronts to approach their asymptotic state. For all of the 2D simulations, the fronts are initiated using a constant temperature hot sidewall at $x\!=\!0$ (red) and the fronts propagate from left to right.

\begin{figure}[h!]
    \begin{center}
        \includegraphics[width=0.9\linewidth]{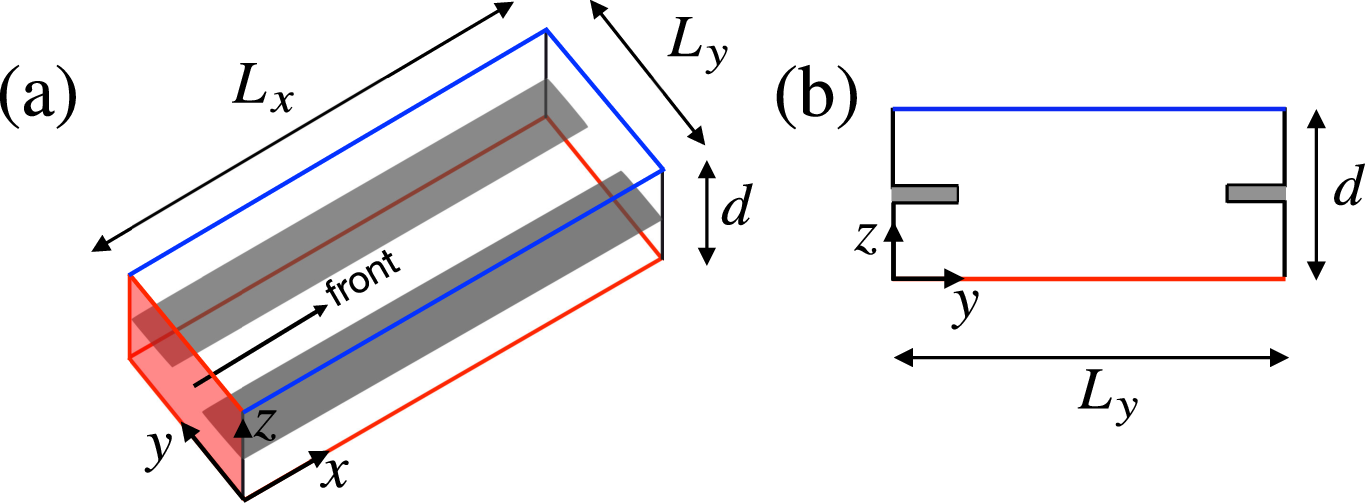}
    \caption{A box domain with fins attached to the sidewalls.  (a) A fluid layer of depth $d$ with a rectangular cross section of lengths $L_x$ and $L_y$ in the $x$ and $y$ directions, respectively to yield $\Gamma_x \!=\! L_x/d$ and $\Gamma_y \!=\! L_y/d$. The bottom wall is hot (red) and the top wall is cold (blue). Thin fins are attached to the sidewalls (shown in gray).  The front is initiated at the wall located at $x \!=\! 0$ (red) and propagates in the positive $x$-direction.  (b)~A $y$–$z$ cross-section highlighting the fin geometry. The fins are no-slip surfaces of finite thickness in the $y$ and $z$-directions and are perfect thermal conductors. We also use a box domain without fins (not shown) which is this domain with the gray fins removed. Gravity opposes the $z$-direction and the schematics are not  drawn to scale.}
    \label{fig:fig2}
    \end{center}
\end{figure}

We also explore the propagation of fronts of straight parallel rolls in 3D box domains, see Fig.~\ref{fig:fig2}. The bottom surface is hot at $T_h$, the top surface is cold $T_c$, and gravity opposes the $z$-direction.  The first box domain explored, not shown explicitly in Fig.~\ref{fig:fig2}, has finite lengths in the $x$ and $y$-directions of  $L_x$ and $L_y$, respectively, to yield aspect ratios $\Gamma_x \!=\! L_x/d$ and $\Gamma_y \!=\! L_y/d$.  We have used $\Gamma_x\!=\!27.3$ and $\Gamma_y \!=\! 6.54$ to align with the apparatus used in the experiments of~\cite{fineberg1987vortex}. For this domain, $\Gamma_x \!\gg\! \Gamma_y$, and as a result the convection rolls orient their axes along the direction of the shorter side of the domain. As a result, straight parallel rolls form with their axes aligned in the $y$-direction and the formation of new rolls propagates in the positive $x$-direction as indicated by the arrow.

We also investigate this box domain with the addition of thin fins attached to the sidewalls, see Fig.~\ref{fig:fig2}, as used in the experiments of~\cite{fineberg1987vortex}. The fins are attached to the sidewalls, at mid-height ($z \!=\! 1/2$), and extend from the sidewalls into the domain. This is shown in Fig.~\ref{fig:fig2}(a) where the fins are the gray surfaces. The wall where the front is initiated, at $x\!=\!0$, is shown in red and the front propagates in the positive $x$-direction.

A $y$-$z$ cross-section of the domain is shown in Fig.~\ref{fig:fig2}(b)  highlighting the fin geometry. The fins have a thickness, in the $z$-direction, of 0.1 and they extend from the sidewall into the domain by a distance of 1 where distance has been nondimensionalized using the layer depth $d$. All material surfaces, including the fins, are no-slip surfaces and the fins are perfect thermal conductors.

In experiment, it is often desired to include a ``soft" boundary~\citep{kramer:1982} on the sidewalls that is less restrictive than a no-slip surface. Examples include ramped and finned boundaries~\citep{daviaud:1989,debruyn:1996,bajaj:1999,paul:2002:pre,paul2003pattern}.  The intention behind using a finned boundary is often to provide a sidewall boundary condition with reduced thermal forcing and reduced viscous shear. The fins are no-slip surfaces extending into the domain that quench convection in the region above and below the fins. The quenching of the fluid motion can be traced to the cubic depth dependence of the Rayleigh number.  However, the influence of the finned boundaries on the convection rolls, and on the propagating front, is quite complex~\citep{paul2003pattern}.  A thorough theoretical understanding of how the fins affect the fluid dynamics is not currently available and we use experimentally accurate numerical simulations to explore this further.  

\begin{figure}[h!]
    \begin{center}
        \includegraphics[width=0.5\linewidth]{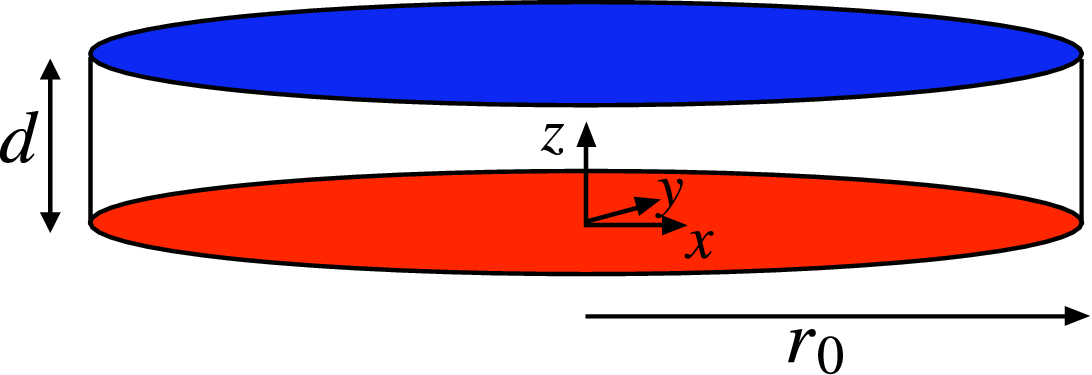}
    \caption{Cylindrical domain used to study the propagation of concentric convection rolls. A fluid layer of depth $d$, radius $r_0$, and aspect ratio $\Gamma \!=\! r_0/d$ with a hot bottom wall (red) at $T_h$ and cold top wall (blue) at $T_c$. All material boundaries are no-slip surfaces. The front is initiated at the origin by introducing a thermal perturbation at $t\!=\!0$. The front on concentric convection rolls propagates radially outward from the origin toward the sidewalls. Gravity opposes the $z$-direction and in our study $\Gamma \!=\! 40$.}
    \label{fig:fig3}
    \end{center}
\end{figure}

Figure~\ref{fig:fig3} shows a cylindrical domain of aspect ratio $\Gamma \!=\! r_0/d$, where $r_0$ is the radius of the domain. The front is initiated at the center ($x \!=\! y \!=\! 0$) by a locally imposed thermal perturbation causing the formation of a concentric convection roll. A concentric convection roll is interesting in light of our study because it represents an experimentally accessible 3D case of a convection roll without a sidewall boundary due to it axisymmetric structure. The concentric roll at the center causes the formation of additional rolls resulting in a front propagating radially outward leaving a field of concentric rolls (or target pattern) in its wake.

\subsection{Computational Approach}

The fluid motion due to Rayleigh--B\'enard convection is described by the nondimensional Boussinesq equations  
\begin{eqnarray}
\mathrm{Pr}^{-1} \! \left( \frac{\partial \vec u}{\partial t} \!+\! \vec u \cdot \vec \nabla \vec u \right )\! &=&\! - \vec \nabla p \!+\! \nabla^2 \vec u \!+\! \mathrm{Ra} T \hat z \!  \label{eq:momentum} \\
\frac{\partial T}{\partial t} + \vec{u} \cdot \vec \nabla T &=& \nabla^2 T \label{eq:energy} \\
\vec{\nabla} \cdot \vec{u} &=& 0 \label{eq:mass}
\end{eqnarray}
which represent the conservation of momentum, energy, and mass, respectively where $\vec{u}$ is the fluid velocity vector, $T$ is the temperature, $p$ is the pressure, $\hat{z}$ is a unit vector in the $z$-direction, and $\mathrm{Pr}$ is the Prandtl number. The nondimensionalization is done in the typical manner using $d$ as the length scale, the vertical diffusion of heat $d^2/\kappa$ as the timescale where $\kappa$ is the thermal diffusivity of the fluid, and $\Delta T$ as the temperature scale. In the following, we will assume that all variables are in nondimensional form.

We integrate Eqs.~\eqref{eq:momentum}-\eqref{eq:mass} using a parallel spectral-element approach~\citep{nek5000,deville2002high} that has been used extensively to study open questions regarding fluid convection~\citep{paul2001power,karimi2012quantifying,paul:2002:pre,xu2016covariant,mukherjee2019velocity,scheel:2013,scheel2006lyapunov,mehrvarzi:2014}.  All material surfaces are no slip, $\vec{u}\!=\!0$, and the temperature of the bottom and the top surfaces are $T(z\!=\!0) \!=\! 1$ and $T(z\!=\!1) \!=\! 0$, respectively. \rev{All sidewalls, other than the sidewall where front initiation occurs in the rectangular domains, and the fins, when included, are perfect thermal conductors. As a result, these sidewalls and the fins are held at a constant temperature for all time given by the thermal conduction profile $T(z) \!=\! 1\!-\!z$.}

In the 2D domain (Fig.~\ref{fig:fig1}) the front is initiated using a hot sidewall $T(x\!=\!0) \!=\! T_0$ where we use $T_0 \!=\! 1$. In the box domains, we initiate the fronts using a constant temperature hot sidewall, as well as, a constant heat flux $q''$ sidewall such that $q''(x\!=\!0) \!=\! q_0''$ where $q_0''$ is a constant.  The heat flux has been nondimensionalized using $k_f \Delta T/d$ where $k_f$ is the thermal conductivity of the fluid. Using Fourier's law of heat conduction this boundary condition can also be expressed as $\partial T / \partial x|_{x=0} \!=\! - q_0''$. We have used $q_0''\!=\!0.014$, which aligns with the wall heating used in the experiments of~\cite{fineberg1987vortex}. We have found that $v_f$ and  $q$ do not vary significantly for  $1\times10^{-3} \lesssim q_0'' \lesssim 1$. 

In the cylindrical domain (Fig.~\ref{fig:fig3}) the front is initiated at the center of the domain by imposing a localized Gaussian perturbation to the temperature field which can be expressed as
\begin{equation}
T(x,y,z,t\!=\!0) = e^{-\frac{(x^2 + y^2)}{\eta^2}}
\label{eq:gaussian}
\end{equation}
where $\eta^2 \!=\! 1/2$. \rev{This yields a rapid, localized drop-off of the initial disturbance.}

\subsection{Simulation Cases}
The simulation cases we use are summarized in Table~\ref{tab:simulations}. Simulation~1 uses the 2D domain (Fig.~\ref{fig:fig1}) with a constant temperature wall boundary condition to initiate the front. We use $\Gamma \!=\! 30$  except for simulations near the onset of convection, $\epsilon \!\lesssim\! 0.1$, where we use $\Gamma \!=\! 60$ to allow for the front to settle towards its asymptotic state. We use both  $\mathrm{Pr} \!=\! 1$ and $\mathrm{Pr} \!=\! 5.373$ for $\mathrm{Ra}_c \!\lesssim \!\mathrm{Ra} \!\leq \!10^4$. For $\mathrm{Ra}{\gtrsim}10^4$, the bulk instability of the fluid layer occurs too rapidly to study propagating fronts using our approach. Simulations~2 and 3 use box domains without, and with, fins respectively. For the box domains we use both constant temperature and constant heat flux boundary conditions to initiate the fronts. Simulation~4 uses the cylindrical domain (Fig.~\ref{fig:fig3}) with $\mathrm{Pr} \!=\! 1$ and $3 \!\times \!10^3 \!\leq \!\mathrm{Ra} \!\leq\! 10^4$ where the front is initiated by a thermal perturbation at the center of the domain.
\setlength{\tabcolsep}{10pt}
\setlength\extrarowheight{5pt}
\begin{table*}
    \centering{
     \begin{tabular}{l l l l l l}
         {\bf Sim.} & {\bf Aspect ratio} & {\bf Domain} & {\bf Ra}&{\bf Pr} & {\bf Initiation}\\
         \hline \hline 
          1 & $\Gamma \!=\!30, 60$ & 2D, Fig.~\ref{fig:fig1} & $1708 \! \leq \! \mathrm{Ra} \! \leq \! 10^4$ &1, 5.373 & $T_0$ \\
         2 & $\Gamma_x\!=\!27.3$, $\Gamma_y\!=\!6.54$ & Box without fins & $1708 \! \leq \! \mathrm{Ra} \! \leq \! 1800$ &1, 5.373 & $T_0$, $q''_0$\\
         3 & $\Gamma_x\!=\!27.3$, $\Gamma_y\!=\!6.54$ & Box with fins, Fig.~\ref{fig:fig2} & $1708 \! \leq \! \mathrm{Ra} \! \leq \! 1800$& 5.373 & $T_0, q''_0$\\
         4 & $\Gamma\!=\!40$ & Cylindrical, Fig.~\ref{fig:fig3} & $3\! \times \!10^3 \! \leq \! \mathrm{Ra} \! \leq \! 10^4$&1 & Gaussian\\
    \end{tabular}}
    \caption{Summary of the simulations including the aspect ratio, domain geometry, Rayleigh number, Prandtl number, and the method to initiate the front. $T_0$ indicates a hot wall with a constant temperature, $q_0''$ is a wall with a constant heat flux, and Gaussian represents a localized Gaussian disturbance of the temperature given by Eq.~\eqref{eq:gaussian}.}
    \label{tab:simulations}
\end{table*}

\section{Results and Discussion}

\subsection{Fronts of Straight Parallel Convection Rolls}

\begin{figure}[h!]
    \begin{center}
        \includegraphics[width=0.8\linewidth]{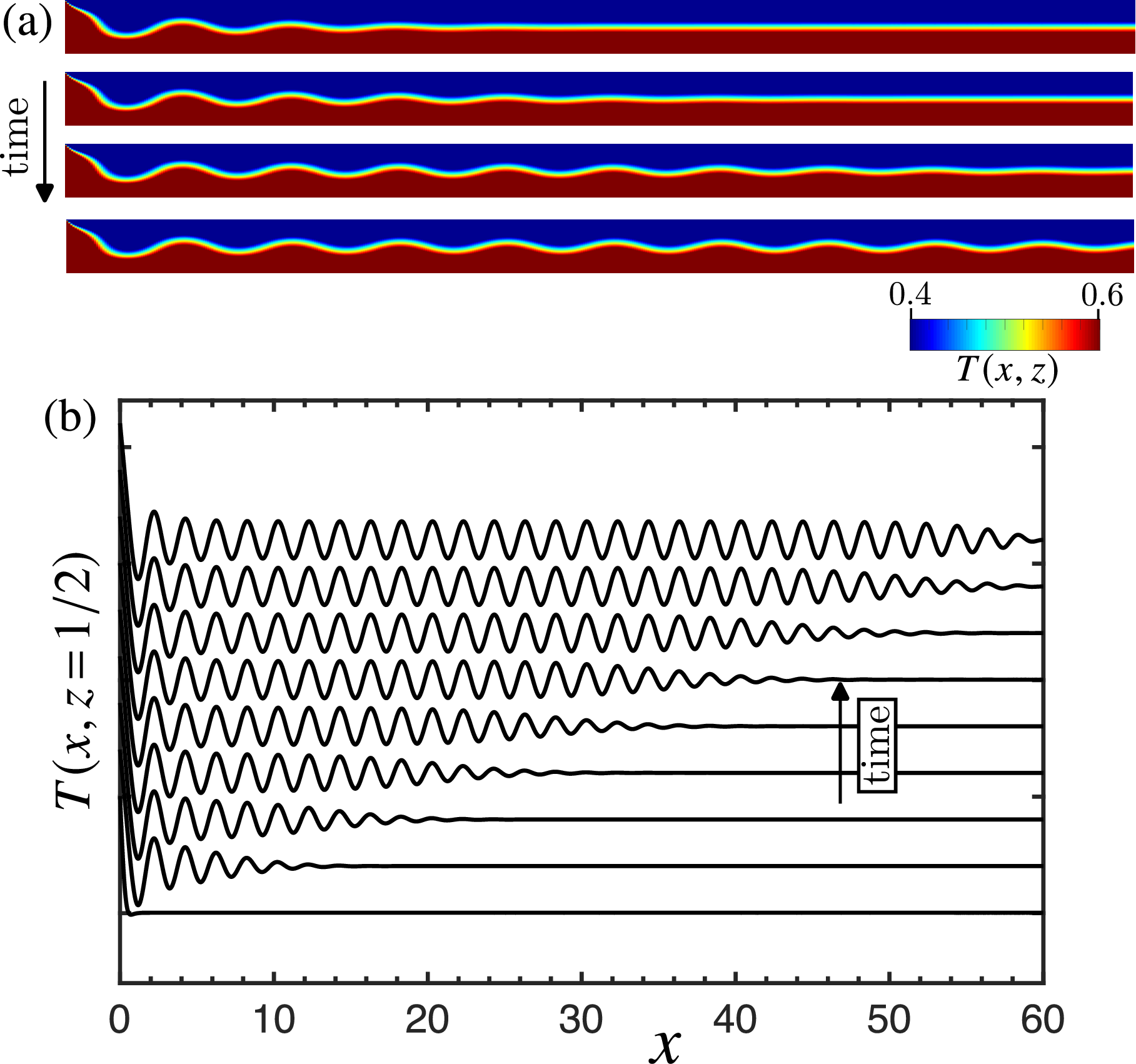}
    \caption{The temperature variation of a fluid layer containing a front in a 2D domain forming a chain of counter-rotating convection rolls. Simulation parameters: Sim.~1, $\Gamma \!=\! 60$, $\mathrm{Ra}\!=\!1745$, $\mathrm{Pr}\!=\!1$. (a) Color contours of $T(x,z)$ at four times, the front propagates from left to right. The region $0 \!\leq \!x \!\leq \! 20$ is shown for clarity. Red is hot rising fluid and blue is cool falling fluid. Time increases from top to bottom. (b) Temperature profiles $T(x,z\!=\!1/2)$ as a function of time $t$. Time increases from bottom to top. The bottom curve is at $t\!=\!0$, the interval between the remaining profiles is 5.68 time units.}
    \label{fig:fig4}
    \end{center}
\end{figure}

We first discuss fronts which leave in their wake a field of straight and parallel convection rolls using the 2D domain (Sim.~1). Figure~\ref{fig:fig4}(a) shows color contours of $T(x,z)$ for $0 \!\leq\! x \!\leq\! 20$ in a domain with aspect ratio $\Gamma\!=\!60$ where $\mathrm{Ra} \!=\! 1745$. The  front is initiated using $T(x\!=\!0)\!=\!1$ and is traveling from left to right. Red and blue contours represent hot and cold fluid, respectively. The temperature field is shown at four instances of time with time increasing from top to bottom.

\begin{figure}[h!]
    \begin{center}
        \includegraphics[width=\linewidth]{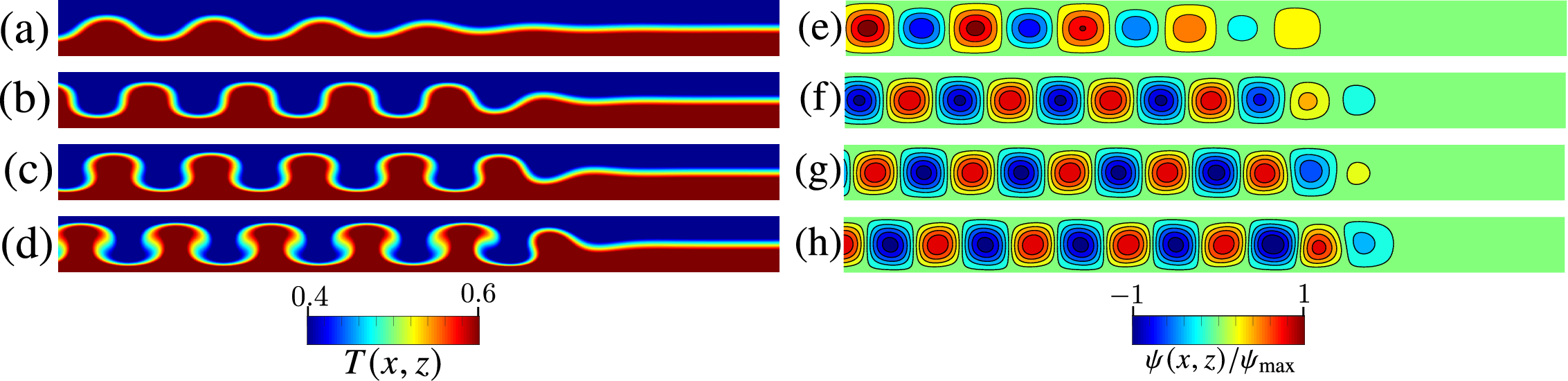}
    \caption{Color contours of temperature (left column) and normalized stream function (right column) for fronts in a 2D domain. Each row is for a different value of $\mathrm{Ra}$ and shows contours at the time when $x_f \!=\! 24$. The spatial region $13 \! \leq \!x \!\leq \!26$ is shown for clarity. (a),(e) $\mathrm{Ra}\!=\!2000$; (b),(f) $\mathrm{Ra}\!=\!3300$; (c),(g) $\mathrm{Ra}\!=\!5000$; (d),(h) $\mathrm{Ra}\!=\!10^4$. Simulation parameters: Sim.~1, $\Gamma \!=\! 30$, $\mathrm{Pr}\!=\!1$.}
    \label{fig:fig5}
    \end{center}
\end{figure}

The temperature field can be used to identify locations with, and without, fluid motion. This is illustrated in Figs.~\ref{fig:fig4}-\ref{fig:fig5} where we show temperature contours, temperature profiles, and contours of the stream function.  In the absence of fluid motion, $T(x,z)$ varies linearly in the $z$ direction due to heat conduction to yield $T \!=\!1 \!-\! z$. This is shown on the far right side of the top panel of Fig.~\ref{fig:fig4}(a) where the fluid is motionless since the front has not yet arrived at this location and, as a result, $T(x,z)$ varies uniformly from red to blue.  However, near the left wall, where convective motion is present,  $T(x,z)$ deviates from its linear variation indicating the presence of hot rising fluid and cool falling fluid due to the convection rolls.

We will use the temperature field at the midplane, $T(x,z\!=\!1/2)$, referred to as the temperature profile, to quantify the position of the front $x_f$ and the wavelength $\lambda$ of the convection rolls. The variation of the temperature profile at several times is shown in Fig.~\ref{fig:fig4}(b). The horizontal line at the bottom is the initial state of no fluid motion, with a linear conduction temperature variation that yields $T(x,z\!=\!1/2)\!=\! 0.5$ over the entire domain. Each curve is the temperature profile at a different instant of time where time increases from the bottom to the top. The different temperature profiles are translated vertically in order to show them clearly on one plot.

The front location $x_f$ is identified as the position where the temperature profile first deviates from a value of 1/2 when examining the temperature profile from right to left (for example, see Fig.~\ref{fig:fig4}(b)). \rev{Numerically, we quantify $x_f$ as the largest value of $x$ such that $|T(x,z\!=\!1/2)\!-\!1/2| \!>\! \delta$ for $\delta \!=\! 1 \times 10^{-3}$. We have verified that our results do not vary significantly upon the value of the threshold used over the range $1 \!\times 10^{-4} \! \leq \! \delta \! \leq \! 1 \! \times \!10^{-2}$.} The front location separates quiescent fluid ($x \!>\! x_f$) from  convective fluid motion ($x \! \le \! x_f$). The front velocity is the time rate of change of the front location, $v_f = \dot{x}_f$.

Peaks, or local maxima, in the temperature profiles shown in Fig.~\ref{fig:fig4}(b) indicate rising hot fluid and troughs, or local minima, indicate descending cold fluid. The $x$ location of the center of a roll pair can be identified as a local maximum in the temperature profile. The local maximum indicates the upflow which occurs at the shared roll  boundary at the center of the roll pair. This roll pair is then bounded on its left, and right, by adjacent local minima  indicating the region of downflow for each roll of the pair. A similar argument can be made using a local minimum in the temperature profile to locate the center of a roll pair where the adjacent local maxima locate the outer edges of the roll pair. The centers of individual convection rolls occur at locations where $T(x,z \!=\!1/2)\!=\! 0.5$ in regions where convective motion is occurring. We will quantify $\lambda$ using the location of the roll centers.

The connection between $T(x,z)$ and $\vec{u}(x,z)$ is shown in Fig.~\ref{fig:fig5} where each row shows $T(x,z$) contours on the left and contours of the stream function $\psi(x,z)$ on the right at the time when $x_f \!=\!24$. The stream function is defined in the usual manner as $u \!=\! \partial \psi/\partial z$ and $w \!=\! - \partial \psi/\partial x$ where $u(x,z)$ and $w(x,z)$ are the  $x$ and $z$ components of the fluid velocity vector, respectively. Red indicates counterclockwise fluid motion, blue indicates clockwise fluid motion, and green represents negligible fluid motion. The contours of $\psi(x,z)$  yield the structure of the counter-rotating convection rolls. Only the spatial region $13 \!\leq \!x \!\leq \!26$ is shown to highlight the region near the front.

Figure~\ref{fig:fig5}(a),(e) shows $T$ and $\psi$ contours, respectively, for $\mathrm{Ra}\!=\!2000$.  As expected for this low value of the Rayleigh number, the temperature exhibits a decaying sinusoidal structure as the front is approached from the left, and the variation of $\psi$ indicates the presence of counter-rotating convection rolls in the wake of the front.

As $\mathrm{Ra}$ increases, the spatial structure of $T$ and $\psi$ deviate from this description. Most striking is the transition of $T(x,z)$ toward a plume-like structure. In Fig.~\ref{fig:fig5}(b)-(d), the variation of the sinusoidal temperature contours toward  mushroom shaped structures is evident.  In addition, the spacing between successive peaks in the temperature profile decreases with increasing $\mathrm{Ra}$ indicating a decrease in the wavelength of the convection rolls. As $\mathrm{Ra}$ is increased, the flow field begins to exhibit asymmetries in the convection rolls near the leading edge of the front as shown on the far right of Fig.~\ref{fig:fig5} (d) and~(h).

\begin{figure}[h!]
    \centering
    \includegraphics[width=0.6\linewidth]{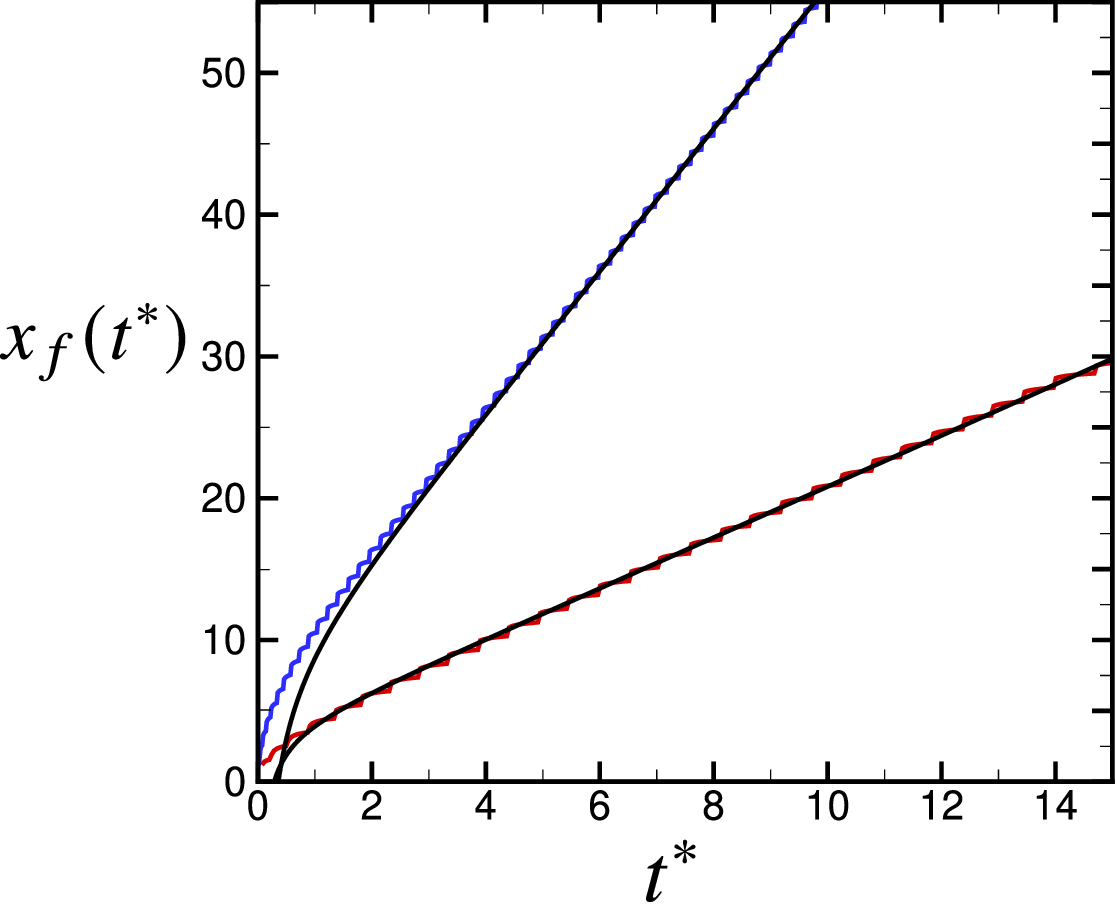}
    \caption{The variation of $x_f$ with $t^*$ for fronts of straight parallel rolls propagating in a 2D domain with $\mathrm{Pr} \!=\! 1$ where $t^* \!=\! \epsilon t/\tau_0$. Upper two curves: $\mathrm{Ra} \!=\! 1745$, $\Gamma \!=\! 60$, numerical results (blue), and the prediction (black) given by Eq.~\eqref{eq:xf-asymptotic}. Lower two curves: $\mathrm{Ra} \!=\! 2000$,  $\Gamma \!=\! 30$, numerical results (red), and the prediction (black) given by Eq.~\eqref{eq:xf-asymptotic}.}
    \label{fig:fig6}
\end{figure}

The variation of $x_f$ with scaled time $t^*$ is shown in Fig.~\ref{fig:fig6} where $t^* \!=\!\epsilon t/\tau_0$ with the characteristic time $\tau_0$ given by $\tau_0^{-1} \!=\! 19.65 \mathrm{Pr}/(\mathrm{Pr} \!+\! 0.5117)$ for RBC with rigid boundaries~\citep{cross1980derivation}. The convergence of $v_f(t)$ to its steady asymptotic value $\bar{v}_f$ is algebraically slow~\citep{ebert:2000}. It has been shown that the convergence is approximately achieved when $t^* \! \gtrsim \! 5$~\citep{kockelkoren2003evidence}. The upper curve (blue) is for $\mathrm{Ra} \!=\! 1745$ with $\Gamma\!=\!60$ and the lower curve (red) is for $\mathrm{Ra} \!=\! 2000$ with $\Gamma\!=\!30$. The slightly jagged nature of the variation of $x_f(t^*)$ is an artifact of how we numerically determine its value from the temperature profiles.

\rev{The Ginzburg--Landau, or amplitude, equation serves as a model system that generalizes properties of complex nonequilibrium systems like RBC near threshold, $\epsilon \to 0$. The amplitude equation is a reduction of the the full Boussinesq equations through an asymptotic expansion in the small parameter $\epsilon$~\citep{newell1969finite,segel1969distant}, and has been compared with experiments and simulations, with very good agreement~\citep{fineberg1987vortex,lucke1987propagating,cross1980derivation}. The equation describing the growth of the finite amplitude, $A(x,t)$, of a slowly varying and spatially periodic state near threshold is:
\begin{equation}
    \tau_0 \frac{\partial A}{\partial t} = \epsilon A + \xi_0^2 \frac{\partial^2 A}{\partial x^2}  - g |A|^2 A.
    \label{eq:amplitude}
\end{equation}
}
A theoretical prediction of the variation of the asymptotic front velocity with $\epsilon$ can be  determined by solving the amplitude equation to yield 
\begin{equation}
    \bar{v}_{f,0} = 2\xi_0 \tau_0^{-1} \epsilon^{1/2}
    \label{eq:vel-theoretical}
\end{equation}
where $\xi_0$ is the correlation length~\citep{van2003front}.  For RBC with rigid boundaries the correlation length is $\xi_0^2 \!=\!0.148$~\citep{cross1980derivation}. The front velocity given by Eq.~\eqref{eq:vel-theoretical} is the \textit{pulled} reaction-diffusion front velocity. Depending on their initiation and the front velocity, propagating fronts can be classified as either pushed or pulled. Pulled fronts propagate with a velocity that is determined by the linearized dynamics at the leading edge of the front. In contrast, \textit{pushed} fronts propagate with a front velocity that is greater than this value and is governed by the nonlinearities behind the leading edge of the front~\citep{van2003front}. 

\rev{It is interesting to note the similarity between autocatalytic reaction-diffusion fronts and the amplitude equation Eq.~\eqref{eq:amplitude}~\citep{van2003front}. The term $\xi_0^2 \partial_x^2 A$ is analogous to diffusion term in an autocatalytic reaction-diffusion equation. The term $\epsilon A - g |A|^2 A$ is analogous to the autocatalytic production term, which when linearized about the base state of $A=0$, yields $\epsilon A$. As in a pulled reaction–diffusion front, the propagation speed is set by the linear spreading rate of the leading edge, which yields the particular form of the front velocity given by Eq.~\eqref{eq:vel-theoretical}.} For linearized dynamics to prevail during the front evolution, the spatial variation of the initial condition must be steeper than $e^{-\sqrt{\epsilon}/\xi_0}$.  We have ensured that this condition is satisfied in our simulations.

The convergence of the front velocity towards its asymptotic value can be expressed as~\citep{kockelkoren2003evidence}
\begin{equation}
    \frac{v_f(t^*)}{\bar{v}_{f,0}} = 1  - \frac{3}{4} t^{*-1} + \frac{3 \sqrt{\pi}}{4} t^{*-3/2} + \mathcal{O}(t^{*-2}).
\label{eq:vel-asymptotic}
\end{equation}
 It will be convenient to integrate Eq.~\eqref{eq:vel-asymptotic} to yield an expression for the front position
\begin{equation}
    x_f(t^*) = \bar{v}_{f,0} \left( t^* - \frac{3}{4} \ln t^* - \frac{3 \sqrt{\pi}}{ 2} t^{*-1/2} \right) + \mathcal{O}(t^{*-1})
\label{eq:xf-asymptotic}
\end{equation}
where we have set the integration constant to zero assuming the front is initially at the origin.

The solid lines (black) in Fig.~\ref{fig:fig6} are the predictions given by Eq.~\eqref{eq:xf-asymptotic}. For $\text{Ra} \!=\! 1745$ (upper two curves) the front position asymptotically converges to the theoretical value for $t^* \!\gtrsim \!5$. For  $\mathrm{Ra} \!=\!2000$ (lower two curves) the asymptotic state is approximately reached for $t^* \gtrsim \! 1.5$. As expected, the time for convergence increases as $\epsilon \!\rightarrow\! 0$.

For all of our 2D results we have ensured that the fronts have significantly approached their asymptotic state prior to quantifying the asymptotic front velocity $\bar{v}_f$. It is important to highlight that in experiment it is often very difficult to reach the asymptotic state due to the large aspect ratio domain that would be required. In this respect, the numerical simulations provide direct access to the long-time asymptotic dynamics. In the experiments of~\cite{fineberg1987vortex} the largest times accessible were  $3 \!\lesssim\! t^* \!\lesssim \!4$.

\begin{figure}[h!]
    \begin{center}
     \includegraphics[width=0.6\linewidth]{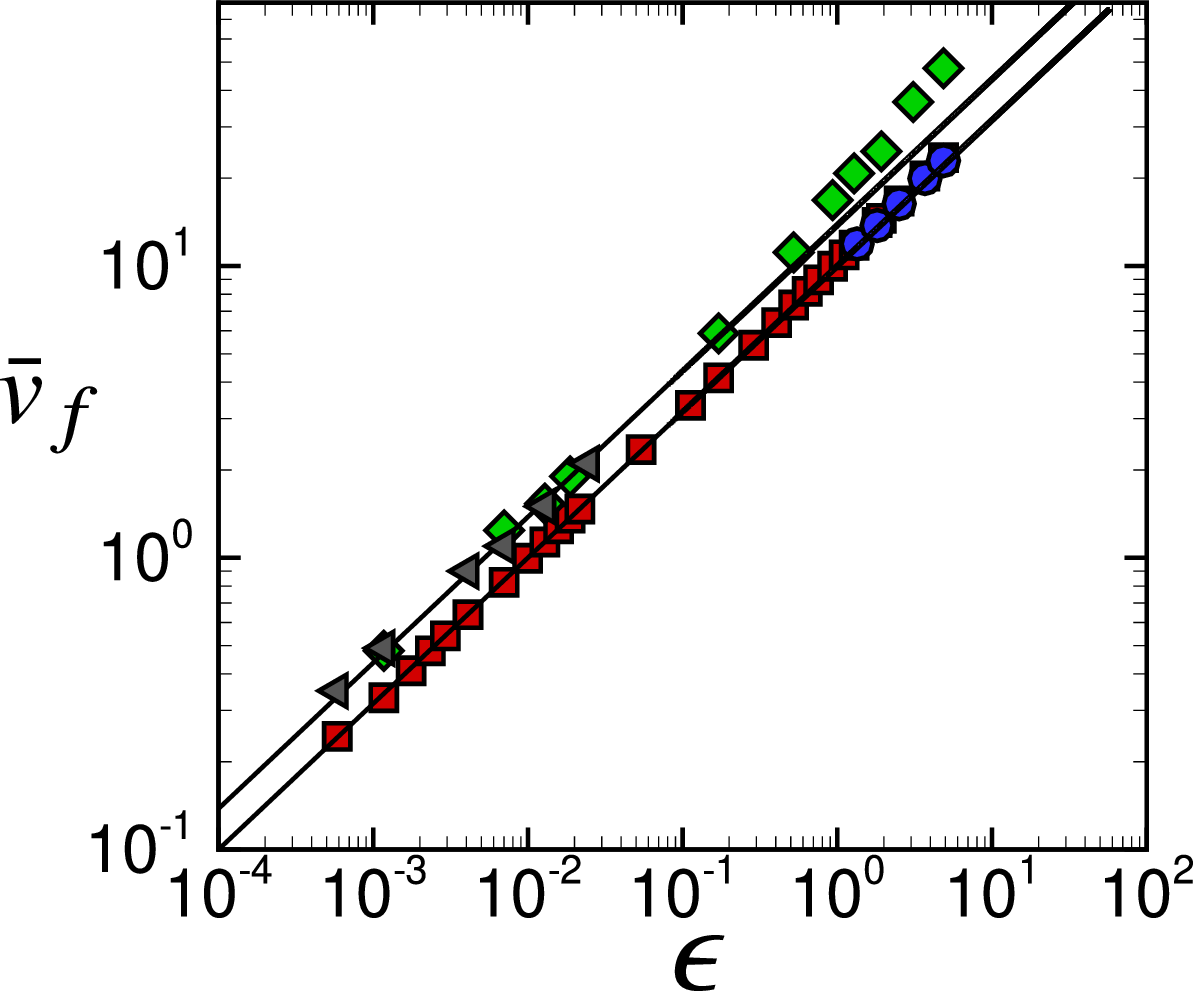}
    \caption{Variation of $\bar{v}_f$ with $\epsilon$, plotted on logarithmic axes: (red squares) 2D domain, $\mathrm{Pr}\!=\!1$; (green diamonds) 2D domain, $\mathrm{Pr}\!=\!5.373$; (gray triangles) box domain with fins, $\mathrm{Pr}\!=\!5.373$;     
    (blue circles) cylindrical domain, $\mathrm{Pr}\!=\!1$. Solid lines are the predicted front velocity using Eq.~\eqref{eq:vel-theoretical}: $\mathrm{Pr}\!=\!1$ (lower),  $\mathrm{Pr}\!=\!5.373$ (upper).}
    \label{fig:fig7}
    \end{center}
\end{figure}

The variation of $\bar{v}_f$ with $\epsilon$ is shown in Fig.~\ref{fig:fig7} for several cases and for an $\epsilon$ variation of over four orders of magnitude. The front velocity for rolls in a 2D domain with $\text{Pr}\!=\!1$ are shown using red squares. The solid lines are the predictions of the front velocity using~Eq.~\eqref{eq:vel-theoretical}. The lower line is the theoretically predicted front velocity with $\text{Pr}\!=\!1$, which yields $\bar{v}_{f,0} \!=\! 10 \epsilon^{1/2}$. The front velocity is well described by the theoretical prediction over the entire range explored. The front velocity for $\text{Pr} \!=\! 5.373$ is shown using green diamonds where the upper line is the predicted front velocity for $\text{Pr} \!=\! 5.373$, which yields $\bar{v}_{f,0} \!=\! 13.8 \epsilon^{1/2}$. For $\epsilon \lesssim 1$ the front velocity follows the $\epsilon^{1/2}$ trend. 

We observe a deviation from the $\epsilon^{1/2}$ trend for $\epsilon \! \gtrsim \! 1$ when Pr\! = \!5.373, where the measured front velocities (green diamonds) exceed the theoretical prediction. The fronts are faster than predicted for a pulled front under these conditions. This deviation could indicate a transition from {pulled} to {pushed} front behavior, where the nonlinearities behind the leading edge increase the front velocity~\citep{van2003front}. \rev{It has been found that adding a symmetry-breaking term in the Swift--Hohenberg model, leads to a transition from pulled to pushed fronts~~\citep{van1988front,van1989front,van2003front}. In the context of RBC, the symmetry breaking could be the transition to the plume-like structure of the flow field, away from the sinusoidal structure  at higher Rayleigh numbers, as shown in Fig.~\ref{fig:fig5}(a)-(d). This transition of the convective rolls occurs at a lower value of the Rayleigh number as the Prandtl number is increased, which could explain why the Pr=1 results follow the theoretical expectations. We, however, do not explore this transition in detail further here.}

Our results show that $\bar{v}_f$ increases with increasing $\text{Pr}$.  The $\bar{v}_f \propto \epsilon^{1/2}$ trend in the front velocity is derived from the amplitude equation in the limit of small $\epsilon$ and it is not expected to be valid for larger values $\epsilon \gtrsim 1$.  It is interesting to note that it continues to describe the fronts well for $\epsilon$ as large as $\epsilon \!\approx\! 10$ for our results with $\text{Pr}\!=\!1$.

\begin{figure}[h!]
    \begin{center}
     \includegraphics[width=0.6\linewidth]{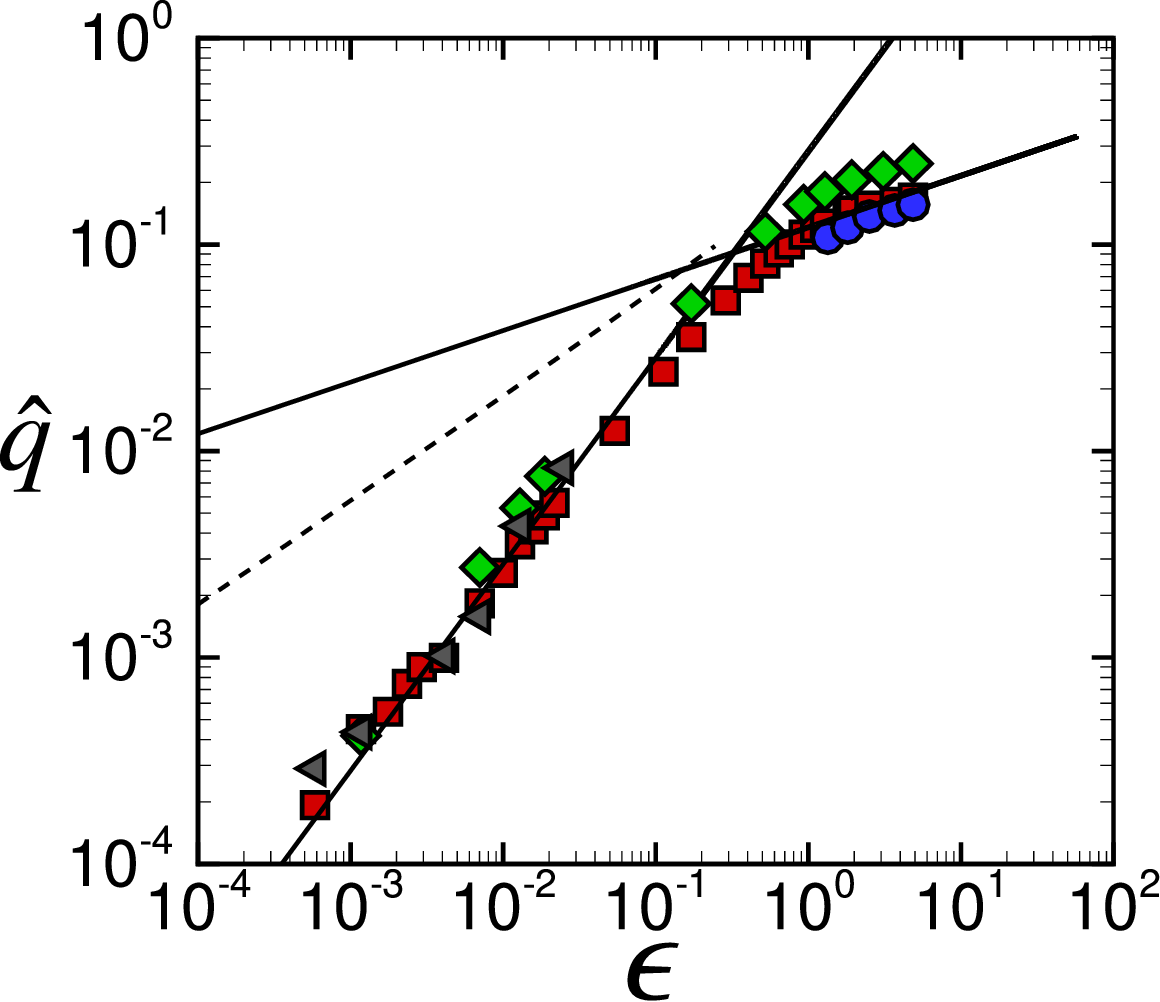}
    \caption{Variation of $\hat{q}$ with $\epsilon$, plotted on logarithmic axes: (red squares) 2D domain, $\mathrm{Pr}\!=\!1$, $q_0\!=\!3.114$; (green diamonds) 2D domain, $\mathrm{Pr}\!=\!5.373$, $q_0\!=\!3.114$; (gray triangles) box domain with fins, $\mathrm{Pr}\!=\!5.373$, $q_0\!=\!3.026$; (blue circles) cylindrical domain, $\mathrm{Pr}\!=\!1$,  $q_0\!=\!3.117$. Solid line (lower), curve fit using $\hat{q} \!=\!0.28 \epsilon$ for $\epsilon \lesssim 0.1$. Solid line (upper), curve fit using $\hat{q} \!=\! 0.12\epsilon^{1/4}$ for $\epsilon \!\gtrsim\! 1$. The dashed line represents the experimental measurements of~\cite{fineberg1987vortex} where $\hat{q} \!\propto\! \epsilon^{1/2}$ with $q_0 \!=\! 2.74$.}
    \label{fig:fig8}
    \end{center}
\end{figure}

The variation of the asymptotic wavenumber with $\epsilon$ is shown in Fig.~\ref{fig:fig8}.  We quantify the time variation of the average wavenumber of the rolls in the following manner. At each time, we determine the wavelengths of the convection rolls that are present as the distance between three adjacent roll centers using the temperature profile. We average the measured wavelengths to obtain an average value of the wavelength for the entire fluid layer at time $t$ which we refer to as $\lambda(t)$. The average wavenumber is $q(t) \!=\! 2 \pi/\lambda(t)$ and we estimate $\bar{q}$ using the fit $q(t) \!=\! \bar{q} + b/t$, where $b$ is a constant. \rev{Both the asymptotic values $\bar{q}$ and $\bar{v}_f$ are obtained from fits to the data which return their values within a 95\% confidence interval. We define the uncertainty as the half-width of this interval, which for the values reported here is within 0.01\% of the fitted values.}

In order to compare the wavenumber trends for multiple cases on one plot, we use the normalized wavenumber $\hat{q} \!=\! (\bar{q}-q_0)/q_0$. We use $q_0$ rather than $q_c$ in this normalization since $q_0 \!\ne\! q_c$ for all of the cases we study. The value of $q_0$ is determined by computing $\bar{q}(\epsilon)$ for decreasing $\epsilon$ and using a linear curve fit to obtain a wavenumber value at $\epsilon\!=\!0$. For our 2D simulations this resulted in $q_0 = 3.114$ (essentially yielding $q_0 \! \approx \! q_c$) where $q_0$ is independent of $\text{Pr}$.

The variation of $\hat{q}$ for the 2D domain, with $\text{Pr} \!=\! 1$, is shown by the red squares.  Near threshold, $\epsilon \lesssim 0.1$, $\hat{q}$ varies linearly  as indicated by the lower solid line.  Further away from threshold, $\epsilon \gtrsim 1$, the variation transitions to a $\epsilon^{1/4}$ dependence as indicated by the fit given by the solid line passing through the results for larger $\epsilon$. The wavenumber variation in the 2D domain with $\text{Pr}=5.373$ is shown by the green diamonds. Overall, the trends are similar with an $\epsilon$ scaling transitioning to $\epsilon^{1/4}$ further from threshold. The wavenumber of the rolls are larger for larger $\text{Pr}$.

The linear scaling of $\hat{q}$ near threshold agrees with the wavenumber $q_m$ which maximizes the growth rate of perturbations in the linear regime~\citep{dominguez1984marginal}.  This can be expressed as $\hat{q}_m \!=\! \alpha \epsilon$ where $q_0\!=\!q_c$. The constant $\alpha$ accounts for the $\text{Pr}$ dependence and is given by
$\alpha \!=\! 0.0494 \!+\! 0.295 \text{Pr} (\text{Pr} \!+\! 0.509)^{-1}$~\citep{dominguez1984marginal}. For $\text{Pr}=1$ this yields $\alpha = 0.245$, for $\text{Pr} \!=\! 5.373$ this yields $\alpha = 0.319$. It is interesting to note that the linear scaling that fits our data, $\hat{q} \!=\! 0.28 \epsilon$, is the mean of these two values of $\alpha$.  A linear scaling of the wavenumber was also reported in a 2D  numerical study for $0.01 \!\le\! \epsilon \!\le\! 0.2$ by~\cite{lucke1987propagating}.

It is insightful to compare these results with the wavenumbers generated by pattern forming fronts using the Swift--Hohenberg equation. In this case  the wavenumber selected behind the front $q_{SH}$ is given by~\citep{van1989front},
\begin{equation}
\frac{q_{SH}}{q_c} = \frac{3 (3+ \sqrt{1 + 6 \epsilon})^{3/2}}{8 (2+\sqrt{1 + 6 \epsilon})}. 
    \label{eq:sh}
\end{equation}
In the limit $\epsilon \! \ll \! 1$ this becomes $\hat{q}_{SH} \!=\! \epsilon/8$ where $q_0\!=\!q_c$ which again recovers the linear variation with $\epsilon$. Additionally, in the limit of $\epsilon \! \gg \! 1$,  Eq.~\eqref{eq:sh} yields $\hat{q}_{SH} \approx 0.587 \epsilon^{1/4}$ in agreement with the large $\epsilon$ trends of the 2D numerical results. \rev{This is expected since the Swift--Hohenberg (SH) equation is a model equation that shares a qualitative structure with the full Boussinesq equations~\citep{swift1977hydrodynamic}.}

The experimental results of~\cite{fineberg1987vortex} are shown in Fig.~\ref{fig:fig8} by the dashed line. The experimentally measured wavenumbers are significantly different than the wavenumbers found in the 2D domain. Furthermore, a striking difference between experiment and the 2D results is the variation of $q_0$. In experiment~\citep{fineberg1987vortex} $q_0\!=\!2.74$ and in our 2D numerics $q_0\!=\!3.114$. The wavelength of the convection rolls at critical in experiment is over 13\% larger than what is found in the 2D numerics. \rev{We re-emphasize that linear theory of bulk convective instability in an infinite fluid layer predicts $q_c=3.117$, whereas the front-selected wavenumber at critical, in both experiments by~\cite{fineberg1987vortex} and our numerical results, satisfy $q_0 < q_c$. Our 2D numerical results yield $q_0=3.114$ which is very close to $q_c$ in magnitude, approximately 0.1\% smaller, while $q_0$ from ~\cite{fineberg1987vortex} is approximately 12\% smaller than $q_c$. In Sim.~3, where we reproduce the experimental domain with finned sidewalls and a constant flux at the left wall, the front-selected wavenumber $q_0$ is still about 2\% less than $q_c$.}

\begin{figure}[h!]
    \begin{center} \includegraphics[width=0.8 \linewidth]{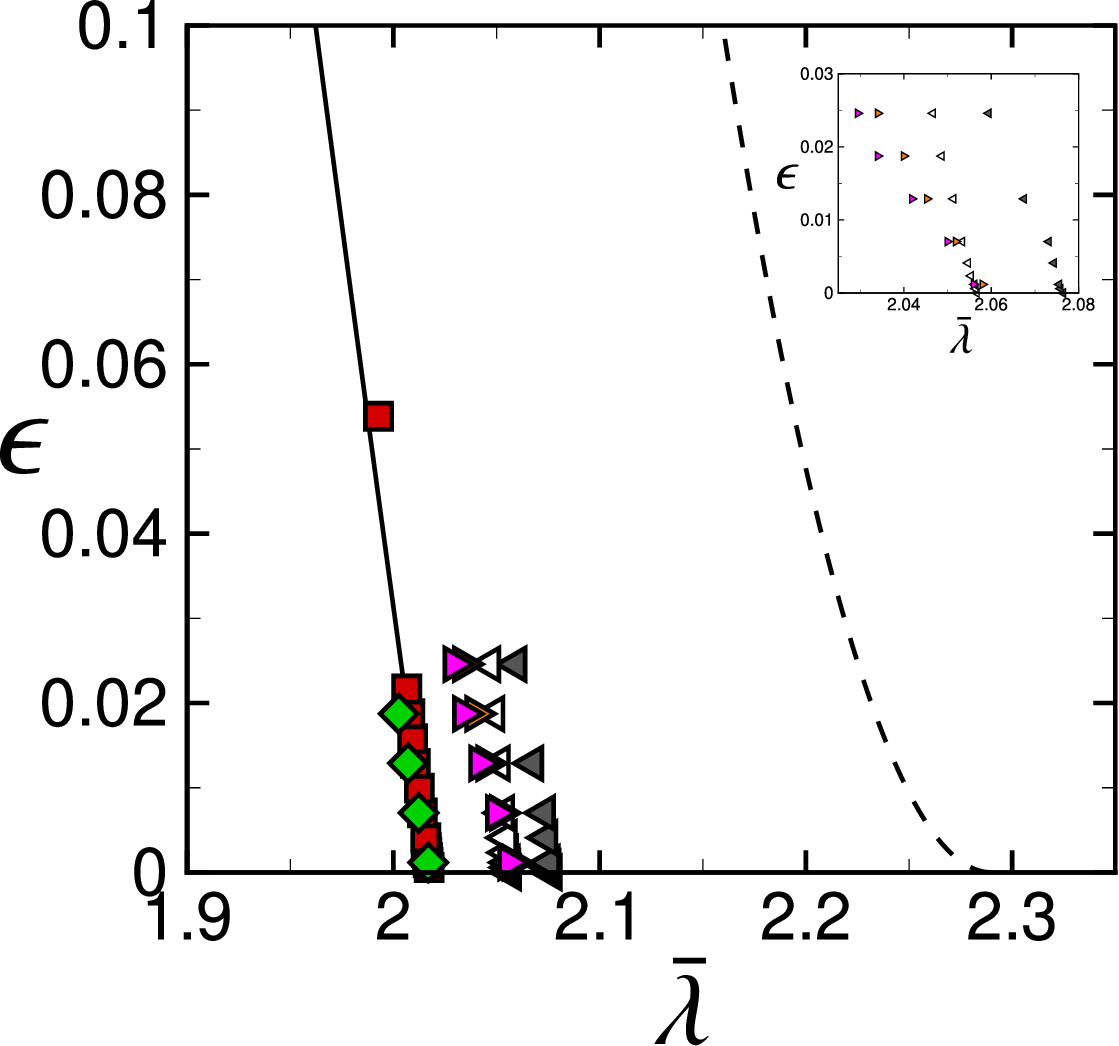}
    \caption{Variation of the wavelength $\bar{\lambda}$ with $\epsilon$: (red squares) 2D domain, $\text{Pr}\!=\!1$; (green diamonds) 2D domain, $\text{Pr}\!=\!5.373$; (pink right triangles) box domain without fins and a left wall with constant temperature $T_0$; $\text{Pr}\!=\!5.373$ for this and all the subsequent cases; (orange right triangles) box domain without fins and a left wall with constant heat flux $q_0''$; (white left triangles) 3D box domain with fins and a left wall held at constant temperature $T_0$; (gray left triangles) box domain with fins and a left wall with constant heat flux $q_0''$. The solid line through the 2D results is $\bar{\lambda}(\epsilon) \!=\! \lambda_0(1+\alpha \epsilon)^{-1}$ where $\lambda_0 \!=\! 2.018$ and $\alpha \!=\! 0.28$. The dashed line denotes the experimental measurements~\citep{fineberg1987vortex} given by $\lambda \!=\! \lambda_0 (1-b\sqrt{\epsilon})$ with $\lambda_0 \!=\! 2.29$, and $b \!=\! 0.18$. In all cases shown, $T_0\!=\!1$ and $q_0''\!=\!0.014$. (inset) A close up view to clearly show the differences in the numerical results for the different 3D cases shown.}
        \label{fig:fig9}
    \end{center}
\end{figure}

We illustrate this more clearly in Fig.~\ref{fig:fig9} where the variation of the asymptotic wavelength with $\epsilon$ is shown near critical. The experimental results of~\cite{fineberg1987vortex} for $\text{Pr} \!=\! 5.373$ are shown by the dashed line. Our 2D numerical results using $\text{Pr}\!=\!5.373$ are the green diamonds. The significant difference in $\lambda_0$ is clearly evident. For reference, we also include the 2D results with $\text{Pr} \!=\! 1$ as the red squares illustrating an insensitivity of $\lambda_0$ on $\text{Pr}$.  

In comparison to the 2D results, the asymptotic wavelength $\bar{\lambda}$ obtained from the simulations using box domains (Sims. 2-3), are larger. We explore box domains with and without fins, and initiate convection with either a constant temperature $T_0$ or a constant flux $q_0''$ at the left wall. For ease of comparison, the inset of Fig.~\ref{fig:fig9} shows a close-up view of the results using the box domains. 

The pink right triangles are for the box domain without fins, where we have used a constant temperature $T_0$ at the left wall to initiate the front. This box domain  can be thought of as a lateral extension of the 2D domain in the $y$ direction (red squares and green diamonds). The convection rolls, in this case, select larger $\bar{\lambda}$ in comparison to 2D. When the same configuration is driven by $q_0''$ on the left wall instead of constant $T_0$, the wavelength of the convection rolls  increases further, as shown by the orange right triangles.

Including fins on the sidewalls of the box domain (Sim.~4) further modifies the asymptotic wavelength. The white left triangles are from the box domain with fins using a constant $T_0$ to initiate the rolls at the left wall. The presence of fins increases the wavelength of the convection rolls when compared with the box domain without fins (pink right triangles). Finally, a box domain with fins using a constant flux initiation $q_0''$ yields the largest values of $\bar{\lambda}$ in our study as shown by the gray left triangles.

\begin{figure}[h!]
    \begin{center}
        \includegraphics[width=0.7\linewidth]{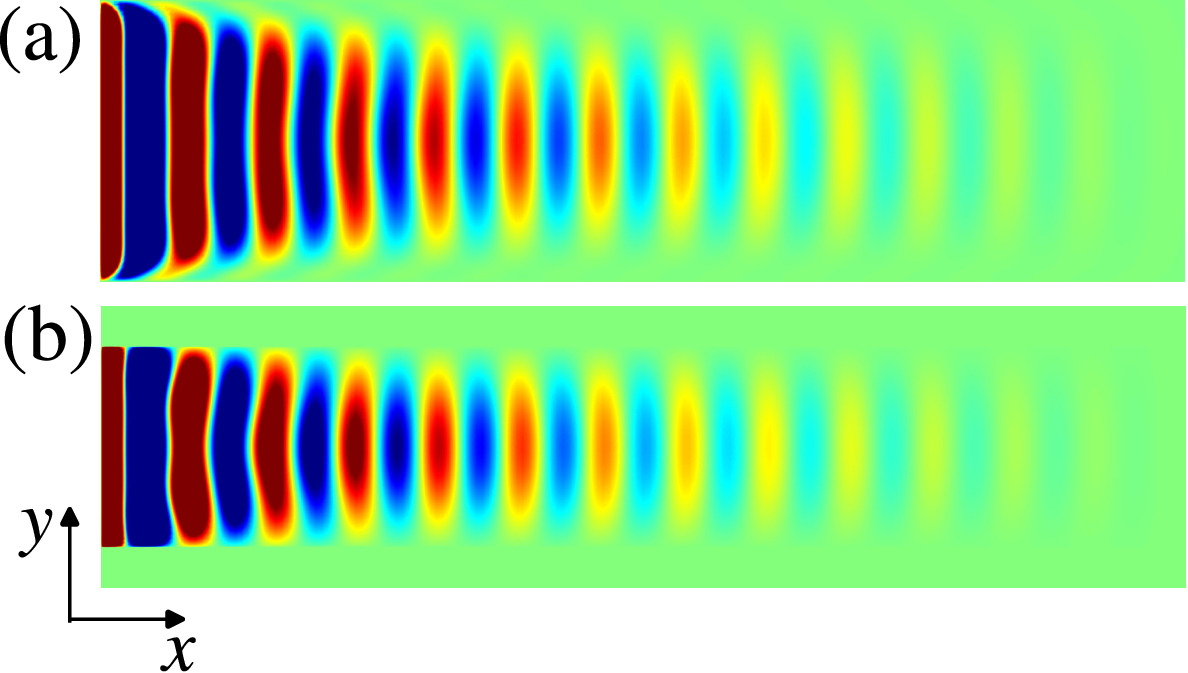}
    \caption{Propagating fronts in box domains ($\mathrm{Ra}\!=\!1710$, $\text{Pr} \!=\! 5.373$).  Contours of the fluid velocity in the $z$-direction at the horizontal midplane, $w(x,y,z\!=\!1/2)$. Red is rising fluid, blue is falling fluid, and green is negligible fluid motion. (a) Box domain (Sim. 3). (b) Box domain with fins (Sim.~4). In both panels the front is initiated using a constant heat flux at $x\!=\!0$ of $q_0''=0.014$.}
    \label{fig:fig10}
    \end{center}
\end{figure}

The influence of adding fins to the box domain is directly reflected in the resulting fluid dynamics.  Flow field images are shown in Fig.~\ref{fig:fig10}(a) using color contours of the fluid velocity in the $z$-direction at the horizontal midplane, $w(x,y,z\!=\!1/2)$. The front is traveling from left to right where red is rising fluid, blue is falling fluid, and green indicates negligible fluid motion. It is clear from these flow fields that the propagating front of convection rolls is a 3D phenomena. For example, the growth of the new convection rolls initially occurs in the center of the domain ($y \! \approx \! \Gamma_y/2$) away from the no slip sidewalls located at $y \!=\! 0$ and $y \!=\! \Gamma_y$.

\begin{figure}[h!]
    \begin{center}
        \includegraphics[width=\linewidth]{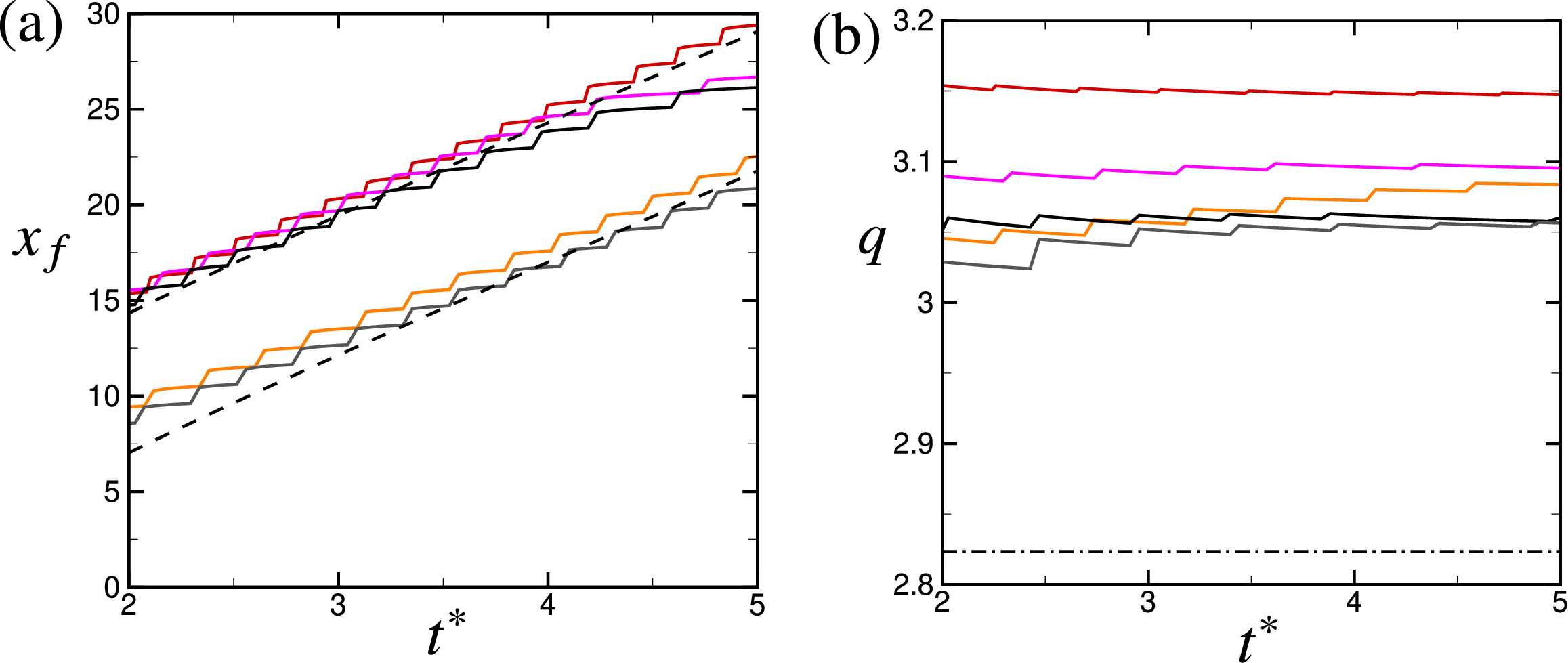}
    \caption{The variation of $x_f$ and $q$ with scaled time for fronts of straight parallel rolls in 2D and box domains ($\mathrm{Ra}\!=\!1750$, $\text{Pr} \!=\! 5.373$). The variation of (a) $x_f(t^*)$ and (b) $q(t^*)$:  (red) 2D domain with constant temperature sidewall; (pink) box domain with constant temperature sidewall; (orange) box domain with constant flux sidewall; (black) finned box domain with constant temperature sidewall; (gray) finned box domain using a constant heat flux sidewall ($q_0'' \!=\! 0.014$). All cases with a hot sidewall use $T_0 \!=\! 1$. The black dashed lines in panel (a) are obtained from Eq.~\eqref{eq:xf-asymptotic} and the dash-dotted line in (b) is $\bar{q}$ from experiment~\citep{fineberg1987vortex}.}
    \label{fig:fig11}
    \end{center}
\end{figure}

The variations of $x_f$ and $q$ with the scaled time $t^*$ are shown by the solid curves in Figs.~\ref{fig:fig11}(a)-(b) for $\mathrm{Ra}\!=\!1750$ and $\mathrm{Pr}\!=\!5.373$. Only the time window $2 \!\lesssim\! t^* \!\lesssim\! 5$ is shown. For $t^* \!\lesssim\! 2$, the front position $x_f$ deviates significantly from the asymptotic state.  For $t^* \!\gtrsim\! 5$, the convection rolls in the box domains interact with the far right boundary of the domain.  

The front position is consistent with the prediction given by Eq.~(\ref{eq:xf-asymptotic}) as shown by the black dashed lines in Fig.~\ref{fig:fig11}(a). The color conventions used here are consistent with Fig.~\ref{fig:fig9}. The red, pink, and black solid curves correspond to the 2D domain, box domain, and box domain with fins, respectively, with each initiated by using a constant temperature $T_0$ at the left wall. The orange and gray curves correspond to the box domain and box domain with fins, respectively, which have been initiated with a constant heat flux $q_0''$. 

It is interesting to note that the front position, when initiated with the constant temperature sidewall, is always larger than the front position when the initiation is done with the constant heat flux sidewall despite each having the same asymptotic front speed. The difference results from a delay in the initiation of the fronts when initiated with a constant flux. The initial heat flux into the fluid for the constant temperature sidewall case is much larger than the value of $q_0''$ that we use. As a result, more time is required for the convection roll initiation in the constant heat flux case. Despite this initial offset, the asymptotic front velocity selected by all the cases remain identical, as indicated by the slopes of the two black dashed lines. We anticipate that this delay in initiation is a function of $q_0''$ and we have not explored this further here. 

The variation of the wavenumber in Fig.~\ref{fig:fig11}(b) is quite interesting for several reasons. The numerical results are shown by the colored curves (using the conventions of panel (a)) and the experimental measurement is the dashed-dotted line. These results indicate that the wavenumber in the 2D domain, with initiation using $T_0$, is the largest (red). A box domain yields a smaller wavenumber (pink) when using $T_0$ to initiate the front. The wavenumber reduces further for a box domain which uses a constant heat flux sidewall for front initiation (orange).  The inclusion of fins to the box domain reduces the wavenumber further,  the black curves use $T_0$ and the gray curve uses $q_0''$ for initiation. Overall, we find that the smallest wavenumber, on average, occurs for the configuration that most aligns with the experiments of~\cite{fineberg1987vortex}. However, the wavenumbers in all of the numerical simulations are significantly larger than those measured experimentally as indicated by the dash-dotted line.

 It is useful to note that for values of $\mathrm{Ra}$ near onset, the accessible time window for the study of front propagation is limited to $t^* \lesssim \!3$ due to the size of the box domains. Therefore, it is expected that the fronts will not have reached their asymptotic states during this time. For these cases, we fit the available data for $q(t)$ and $x_f(t)$ to determine the asymptotic values of wavenumber and front speed.

The front velocity in the box domain with fins is similar to what is found using a 2D domain. This is illustrated by comparing the green diamonds and the gray triangles in Fig.~\ref{fig:fig7}. Similarly, the variation of the scaled wavenumber of the rolls in the box domain with fins is similar to the results from the 2D domain as illustrated in Fig.~\ref{fig:fig8}. However, the wavelength of the rolls at critical is larger for the box domain with fins as indicated in Fig.~\ref{fig:fig9}. Although the wavelength has increased, it still remains smaller than the experimental values indicated by the dash-dotted line in Fig.~\ref{fig:fig11}(b).

\subsection{Fronts of Concentric Rolls}

\begin{figure}[h!]
    \begin{center}
        \includegraphics[width=0.9\linewidth]{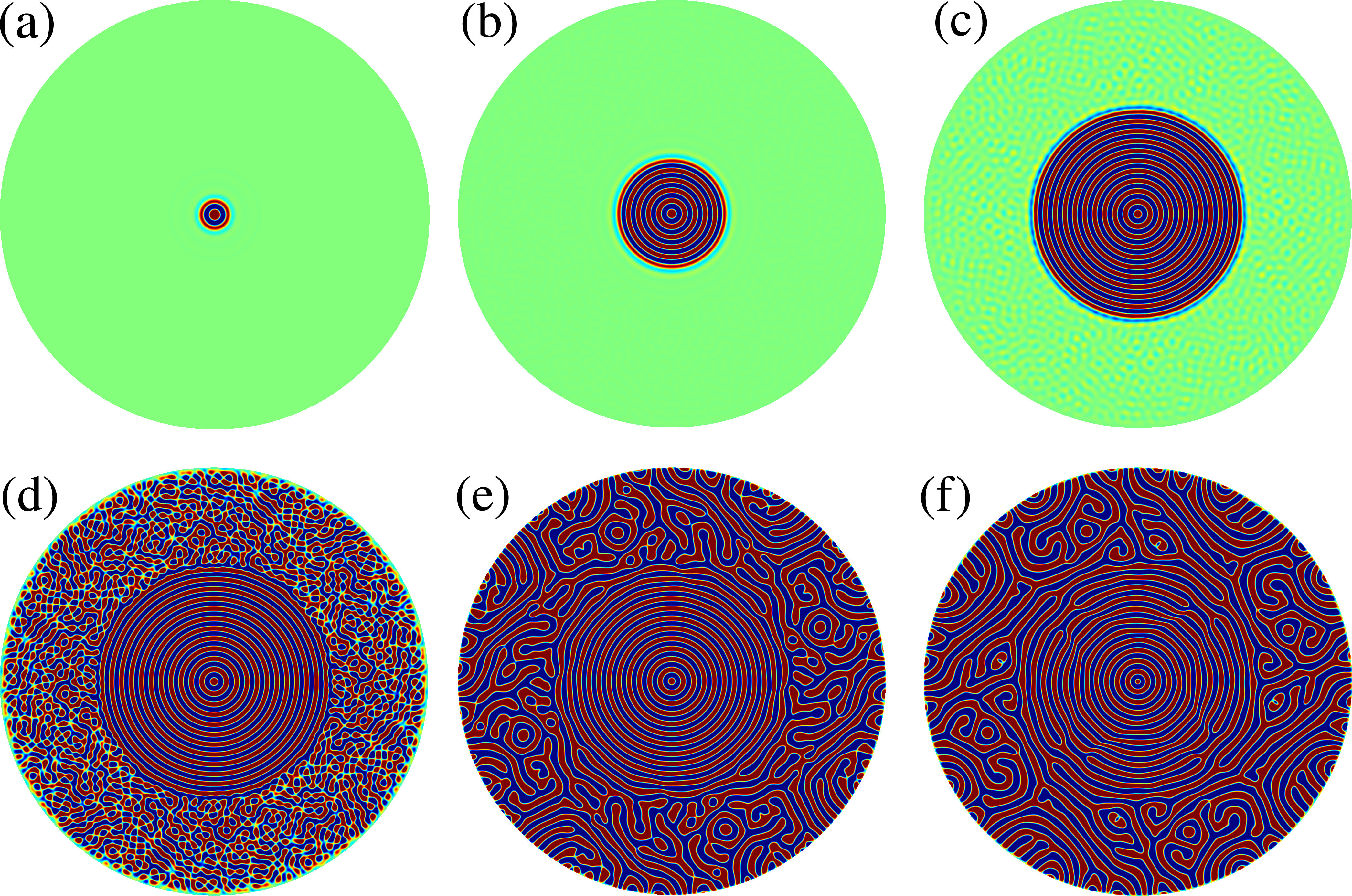}
    \caption{The propagation of concentric convection rolls in a large cylindrical domain: $\mathrm{Ra} \!=\! 3000$ ($\epsilon \!=\! 0.76$), $\mathrm{Pr}\!=\!1$, $\Gamma \!=\! 40$, Sim.~4. The front is initiated at the center of the domain and propagates radially outward towards the sidewalls. Color contours of $T(x,y,z\!=\!1/2)$ are shown at the times: (a)~$t\!=\!0.28$, (b)~$t\!=\!1.2$, (c)~$t\!=\!2.3$, (d)~$t\!=\!2.73$, (e)~$t\!=\!4.62$, (f)~$t\!=\!11$. Red is hot rising fluid and blue is cool falling fluid. For these conditions,  the fluid layer is unstable to the spiral defect chaos state which appears at later times (see panels (d)-(f)).}
    \label{fig:fig12}
    \end{center}
\end{figure}

We next discuss the propagation of a front which forms concentric convection rolls in a cylindrical domain (Sim.~4).  Figure~\ref{fig:fig12} shows color contours of the temperature field at the horizontal midplane, $T(x,y,z\!=\!1/2)$, for a front at several instances of time. The front propagates radially outward towards the boundary as seen in Fig.~\ref{fig:fig12}(a)-(c). At later times, shown in Fig.~\ref{fig:fig12}(d)-(f), the bulk convective instability leads to the emergence of the spiral defect chaos state~\citep{morris1993spiral,bodenschatz2000recent,vitral2020spiral} in the region beyond the front. For very long times (not shown) the concentric rolls are annihilated and the entire fluid layer exhibits spiral defect chaos.

The velocity of the front leaving concentric rolls in its wake is shown in Fig.~\ref{fig:fig7} by the blue circles. The front velocity again follows the $\epsilon^{1/2}$ scaling as predicted by Eq.~\eqref{eq:vel-theoretical}. The scaled wavenumber of the concentric rolls are shown in Fig.~\ref{fig:fig8} by the blue circles where we find $\hat{q} \!=\! 0.12 \epsilon^{1/4}$ in agreement the $\epsilon^{1/4}$ scaling predicted by the Swift--Hohenberg equation away from threshold. It is interesting to point out that the wavelength of the propagating concentric rolls is smaller than the wavelength of the disordered rolls present in the spiral defect region (see Fig.~\ref{fig:fig12}(e) and~(f)). \rev{It is useful to note that the concentric convection roll front is initiated with the perturbation given by Eq.~\eqref{eq:gaussian} which has been chosen to ensure a rapid, localized drop-off of the initial disturbance. We have not evaluated the dependence of the selected wavenumbers against the details of this perturbation, this would be an interesting direction of future work.}

It is insightful to compare our results for the wavenumber of concentric rolls formed behind a front with the wavenumber selected by concentric rolls in the long-time limit due to the bulk instability. The asymptotic wavenumber selected by concentric rolls which completely fill a cylindrical domain has been studied experimentally and theoretically in detail~\citep{cross1993pattern,bodenschatz2000recent}. It has been shown that the asymptotic wavenumber selected by concentric rolls due to the bulk instability \emph{decreases} with increasing $\epsilon$ for $\text{Pr} \gtrsim 1$~\citep{buell1986wavenumber}.  This must be contrasted with the findings for the concentric rolls that form behind a front which yield an increase in the selected wavenumber with increasing $\epsilon$ as indicated in Fig.~\ref{fig:fig8} and the $\hat{q} \!\sim\! \epsilon^{1/4}$ scaling that is shown. The values of the wavenumbers are also significantly different,  the concentric rolls from the bulk instability yield much smaller wavenumbers than what is found for the front selected wavenumbers. For example, the wavenumber selected by the concentric convection rolls shown in Fig.~\ref{fig:fig12} is $\bar{q}\!=\!3.47$. However, for these conditions ($\text{Ra}\!=\!3000$, $\text{Pr}\!=\!1$) the bulk instability selects a wavenumber of $\bar{q} \!\approx \!2.82$~\citep{buell1986wavenumber}.  The average selected wavenumber for the spiral defect chaos state is even smaller with a value of $\bar{q} \!\approx \!2.5$~\citep{chiam:2003}.

In a recent study, we explored the formation of convection rolls behind a reaction front that added heat to the fluid while also changing its density~\citep{mukherjee:2022}. The propagating reaction front temporarily annihilated the spiral defect chaos state as it traveled and in its wake a front forming concentric convection rolls emerged (see Figs.~(15)-(16) in~\cite{mukherjee:2022}). 
The wavenumber of the forming concentric rolls was larger than the average wavenumber of convection rolls undergoing spiral defect chaos in the bulk. The average wavenumber selected by the forming convection rolls in the wake of the reaction front for $\mathrm{Ra}\!=\! 6 \times\! 10^3$ was $ \bar{q} \approx 3.9$. This value of the wavenumber is comparable to the wavenumbers we find here for concentric rolls that remain behind a propagating front in an initially quiescent fluid domain.

\section{Conclusion}
We have explored propagating fronts that form convection rolls in Rayleigh--B\'enard convection across a wide range of conditions.  When the Rayleigh number is just above the convective instability, a heated sidewall is used to initiate the propagation of a chain of convection rolls. The pattern forming front selects a unique wavenumber and front velocity that depend on the Rayleigh number, Prandtl number,  and on the details of the convection domain. 

We have quantified the front velocity by tracking the leading edge of the temperature profile. We find that $v_f \!\sim\! \epsilon^{1/2}$ for most of the cases we have explored. The scaling agrees with the theoretical expression for the velocity of pulled fronts. Deviations from the square-root scaling are observed only for $\mathrm{Pr}\!=\!5.373$ when $\epsilon \!\gtrsim \!1$ which we anticipate is due to the increasing role of nonlinearities as $\epsilon$ increases. 

We have explored the slow asymptotic convergence of the front velocity with time. Near onset, the fronts must evolve for a sufficient time such that $t^* \!\gtrsim\! 5$. The aspect ratio of the domain must be tailored with this slow convergence in mind. If a domain is not large enough, the front will make contact with the far sidewall prior to reaching its asymptotic state.

Near the convective threshold, the asymptotic wavenumber scales linearly with the reduced Rayleigh number. This is in agreement with the theoretical predictions based on the wavenumber of the maximum growth in the linear regime. For larger $\epsilon$ the wavenumber transitions to a $\epsilon^{1/4}$ scaling which is predicted by an analysis using the Swift--Hohenberg equation.

We have compared our numerical results with the experimental measurements of~\cite{fineberg1987vortex} which reported a $\epsilon^{1/2}$ dependence of the wavenumber near onset. An important comparison is with our Simulation~3 which uses a convection domain constructed to match the geometric details and boundary conditions of the experiment. We find that the wavenumbers selected in the simulations exhibit a linear scaling with $\epsilon$ near threshold.  

We find that the wavenumber selection is influenced by the geometry of the convection domain, boundary conditions, and method of front initiation. In particular, box domains produce rolls with larger wavelengths than what is found using a 2D domain. Similarly, initiating convection using a constant heat flux boundary condition results in larger wavelength rolls than when using a constant-temperature boundary to initiate the front. Additionally, incorporating fins on the sidewalls of the box domains result in convection rolls with larger wavelength.

Simulations designed to incorporate details of the experimental setup of~\cite{fineberg1987vortex} produced rolls with the largest wavelengths among all of the cases we explored, yet the wavelengths remained $\!\sim\!10\%$ smaller than the experimentally measured values. The front velocity $v_f$ and the scaling behavior of the normalized wavenumber $\hat{q}$ with $\epsilon$ do not depend significantly upon the domain geometry, boundary conditions, or method of front initiation that are used. \rev{We, however, note that this universality concerns the scaling exponents of the normalized wavenumbers. The prefactors in the relations and the absolute value of the selected wavenumbers vary appreciably between cases, as shown in Fig.~\ref{fig:fig9}.}

The physical origin of the large wavelength convection rolls found in the experiments of~\cite{fineberg1987vortex} for $\epsilon \!\ll\! 1$ remains an open question. Our results suggest this is not due to the use of a specific geometry, the use of finned sidewalls, or the constant heat flux sidewall that was used to initiate the fronts. \rev{A factor that could contribute to the discrepancy is the finite thermal conductivity of the sidewalls and fins that are used in the experiment, which we include as perfect thermal conductors. Quantifying the role of bounding surfaces with finite thermal conductivity on the wavenumber selection would be an interesting direction in the future.}

\section{Acknowledgments}
We acknowledge many fruitful interactions with Paul Fischer and the Nek5000 user group. SM acknowledges Bibhas Kumar for help with mesh generation. Portions of the numerical computations were conducted with generous support from the Advanced Research Computing center at Virginia Tech and the High Performance Computing center at Iowa State University.

\section*{Declaration of interests}
The authors report no conflict of interest.

\clearpage

\bibliographystyle{plainnat}

\end{document}